\documentclass[12pt]{article}
\usepackage{epsfig}
\newcommand{\be}{\begin{equation}}
\newcommand{\en}{\end{equation}}
\newcommand{\eqa}{\begin{eqnarray}}
\newcommand{\ena}{\end{eqnarray}}

\newcommand{\tr}{\mbox{Tr}}
\newcommand{\noi}{\noindent}

\newcommand{\vzero}{{\vec 0}}

\newcommand{\vx}{{\vec x}}

\newcommand{\vR}{{\vec R}}

\begin{document}

\renewcommand{\theequation}{\arabic{section}.\arabic{equation}}
\renewcommand{\thesection}{\arabic{section}}

\title{\vspace{-2cm} \hfill {\small ITEP-LAT/2003-26} \\ 
\vspace{1cm}
On the gauge dependence of the singlet and adjoint potentials }

\author{V.A. Belavin$^{\rm a}$, V.G.~Bornyakov$^{\rm a,b}$,
V.K. Mitrjushkin$^{\rm a,c}$ \\
$^{\rm a}$ {\small\it Institute of Theoretical and  Experimental
Physics,}\\
{\small\it B.Cheremushkinskaya 25, Moscow, 117259, Russia}\\
$^{\rm b}$ {\small\it Institute for High Energy Physics, Protvino 142284,
Russia}\\
$^{\rm c}$ {\small\it Joint Institute for Nuclear Research, 141980 Dubna, Russia
}\\}

\maketitle

\begin{abstract}

We study gauge dependence of the recently suggested definition of the
singlet and adjoint potentials in $SU(2)$ lattice gauge theory. We find
that in the (time local) maximal tree axial gauge the singlet potential
obtained from the gauge dependent correlator $\mbox{Tr}
L(x)L^\dagger(y)$ differs from that computed in the Coulomb gauge. In
the generalized Coulomb gauge we find the range of the parameter values
in which the singlet potential differs from that in the Coulomb gauge.

\end{abstract}

\section{Introduction} \setcounter{equation}{0}

The free energy of a static quark--antiquark pair -- static quark
potential -- is of great importance for the understanding of the
confinement--deconfinement transition, long distance properties of
chromoplasma and heavy quark phenomenology at finite temperatures. One
expects that the static quark  potential depends on the color channel
one chooses, e.g. singlet or adjoint, and that these potentials are
gauge invariant \cite{mlsv,Nadkarni:1986as}. It is well known that the
correlator of the Polyakov loop gives rise to the gauge invariant
definition of the color averaged potential $~V_{av}(\vR)~$ \cite{mlsv}.
However, there is a problem with the gauge invariant definition of the
singlet $~V_{sing}(\vR)~$ and adjoint $~V_{adj}(\vR)~$ potentials.
Recently, it has been argued \cite{Philipsen:2002az} that it is
possible to overcome this problem and to arrive at a gauge invariant
definition of the singlet and adjoint potentials using the so called
dressed Polyakov lines. The dressing of the source may be viewed as a
gauge transformation, and this approach is equivalent to the definition
of the singlet and adjoint potentials in a certain gauge. It has been
claimed \cite{Philipsen:2002az} that one can use any unique gauge that
is local in time, i.e. the gauge that does not change the spectrum of
the transfer matrix (contrary to the case of the Lorentz/Landau gauge
used in \cite{Nadkarni:1986as,akpsw}). Therefore, to find
$~V_{sing}(\vR)~$ one must calculate the average $~\langle \tr \Bigl(
L(\vR) L^{\dagger}(\vzero)\Bigr)\rangle~$ in the chosen (time local)
gauge, $~L(\vx)~$ being the Polyakov line defined in the next section.
This approach has been used to calculate $~V_{sing}(\vR)~$ and
$~V_{adj}(\vR)~$ in $3D$ \cite{Philipsen:2002az} and $4d$
\cite{Digal:2003jc} $SU(2)$, and in $4d$ $SU(3)$
\cite{Kaczmarek:2002mc}  lattice gauge theories.

The main goal of this note is to study more closely the dependence of
$~V_{sing}(\vR)~$ and  $~V_{adj}(\vR)~$ on the choice of the gauge. To
this purpose we calculated the gauge invariant correlator $~\langle \tr
\,L(\vR) \cdot\tr\, L^{\dagger}(\vzero)\rangle~$ as well as the
averages $~\langle \tr \Bigl( L(\vR) L^{\dagger}(\vzero)\Bigr)\rangle~$
in various local in time gauges : maximal tree axial gauge (AG),
Coulomb gauge (CG) and generalized Coulomb gauge (GCG).

All our calculations were made in $4d$ $SU(2)$ lattice gauge theory in
the finite volume with periodic boundary conditions. Main definitions
and computational details are given in the next section. In section 3
we present our results and make conclusions.

\section{Potentials, gauges and details of simulations}\setcounter{equation}{0}

Let $~N_{\tau}~$ and $~N_{\sigma}~$ be the number of sites along the
timelike and spacelike directions, respectively, and $~a~$ be the
lattice spacing. We choose $~0\le x_i\le N_{\sigma}a-a~$ ($i=1,2,3$)
and $~0\le x_4\le N_{\tau}a-a~$.

In the case of $SU(2)$ lattice gauge theory the color
averaged potential $~V_{av}(\vR)~$ is given by \cite{mlsv}

\be
e^{-V_{av}(\vR)/T} = \frac{1}{4} \,\langle \tr \,L(\vR) \cdot\tr\,
L^{\dagger}(\vzero)\rangle~.
\label{cav}
\en

\noi where $T=1/aN_{\tau}$ is the temperature, $L(\vec x)$ is the
Polyakov line defined as $L(\vec x) = \frac{1}{2} {\rm Tr}
\prod_{x_0=0}^{N_\tau a-a} U_{x_0,\vec{x},0}$, $U_{x\mu}$ is the
lattice field. Following \cite{Nadkarni:1986as} one can introduce the
color singlet and adjoint potentials as follows

\eqa
e^{-V_{sing}(\vR)/T} &=& \frac{1}{2} \,\langle \tr \Bigl( L(\vR) \,
L^{\dagger}(\vzero)\Bigr)\rangle~;
           \label{singl}
\\
\nonumber \\
\frac{3}{4}\, e^{-V_{adj}(\vR)/T} &=& e^{-V_{av}(\vR)/T}
- \frac{1}{4}\, e^{-V_{sing}(\vR)/T}~.
           \label{adjo}
\ena

\noi Eq.'s (\ref{singl}),(\ref{adjo}) have meaning only when some gauge
fixing condition is imposed. Without gauge fixing $~V_{sing}(\vR) =
V_{adj}(\vR) = V_{av}(\vR)$. In this work we employ the following local
in time  gauges.

\begin{itemize}

\item $3D$ maximal tree axial gauge.

In this case the gauge is fixed at every time slice independently
as follows :

\begin{itemize}

\item first we fix $~U_{x1}=1~$ for all
$~x_1,x_2,x_3~$ apart from $~x_1=N_{\sigma}a-a~$;

\item then we fix $~U_{x2}=1~$ for $~x_1=N_{\sigma}a-a~$ and all
$~x_2,x_3~$ apart from $~x_2=N_{\sigma}a-a~$;

\item finally we fix $~U_{x3}=1~$ for $~x_1=x_2=N_{\sigma}a-a~$ and all
$~x_3~$ apart from $~x_3=N_{\sigma}a-a~$.

\end{itemize}

\item Generalized Coulomb gauge.

To fix this gauge one should find the maximum of the functional

\be
F_U^{\lambda}(\Omega) = \sum_x \frac{1}{2} \tr\, U^{\Omega}_{x1}
+ \lambda \sum_x \left[ \frac{1}{2} \tr\, U^{\Omega}_{x2}
+ \frac{1}{2} \tr\, U^{\Omega}_{x3}\right]~,
\label{funcGCG}
\en

\noi with respect to gauge transformations $~\Omega~$, where
$~U^{\Omega}_{x\mu}=\Omega_xU_{x\mu}\Omega^{\dagger}_{x+\mu}~$ and
$~\lambda~$ is some parameter taking values between zero and unity.
Evidently, at $~\lambda=1~$ one arrives at the standard Coulomb gauge.

\end{itemize}

Our main simulations were made on $6\times 18^3$ lattice at
$\beta=2.35$. At this $\beta$
$T/T_c=\sqrt{\sigma(\beta_c)/\sigma(2.35)}=0.76(1)$, where $\sigma$ is
string tension, and $~\beta_c=2.43$ \cite{Engels:1992fs}\footnote{We
took $\sigma(2.35)=0.311(2)$ from \cite{Lucini:2001ej} and
$\sigma(2.43)$ was estimated by interpolation of the data for string
tension taken from the literature.}. 5,000 configurations, separated by
100 combined sweeps consisting of one heat bath and two overrelaxation
sweeps, were collected. To compare our results for the Coulomb gauge
with those of ref. \cite{Digal:2003jc} we also made simulations on
$~4\times 16^3$ lattice at $\beta=2.234$ ($T/T_c=0.80$\,
\cite{Digal:2003jc}). To fix the Coulomb and the generalized Coulomb
gauges we used the ordinary procedure alternating the relaxation and
overrelaxation sweeps.

Statistical errors are calculated by the blocked jackknife method.
To see the dependence of the potentials on the choice of the Gribov copy in
the Coulomb gauge we generated $5$ random gauge copies for every
equilibrium configuration at $\beta=2.234$. No sizable effect of the
Gribov copies has been found.

\section{Results in various gauges}
\setcounter{equation}{0}

\begin{figure}[pt]
\begin{center}
\leavevmode
\hbox{
\epsfysize=11cm
\epsfxsize=14cm
\epsfbox{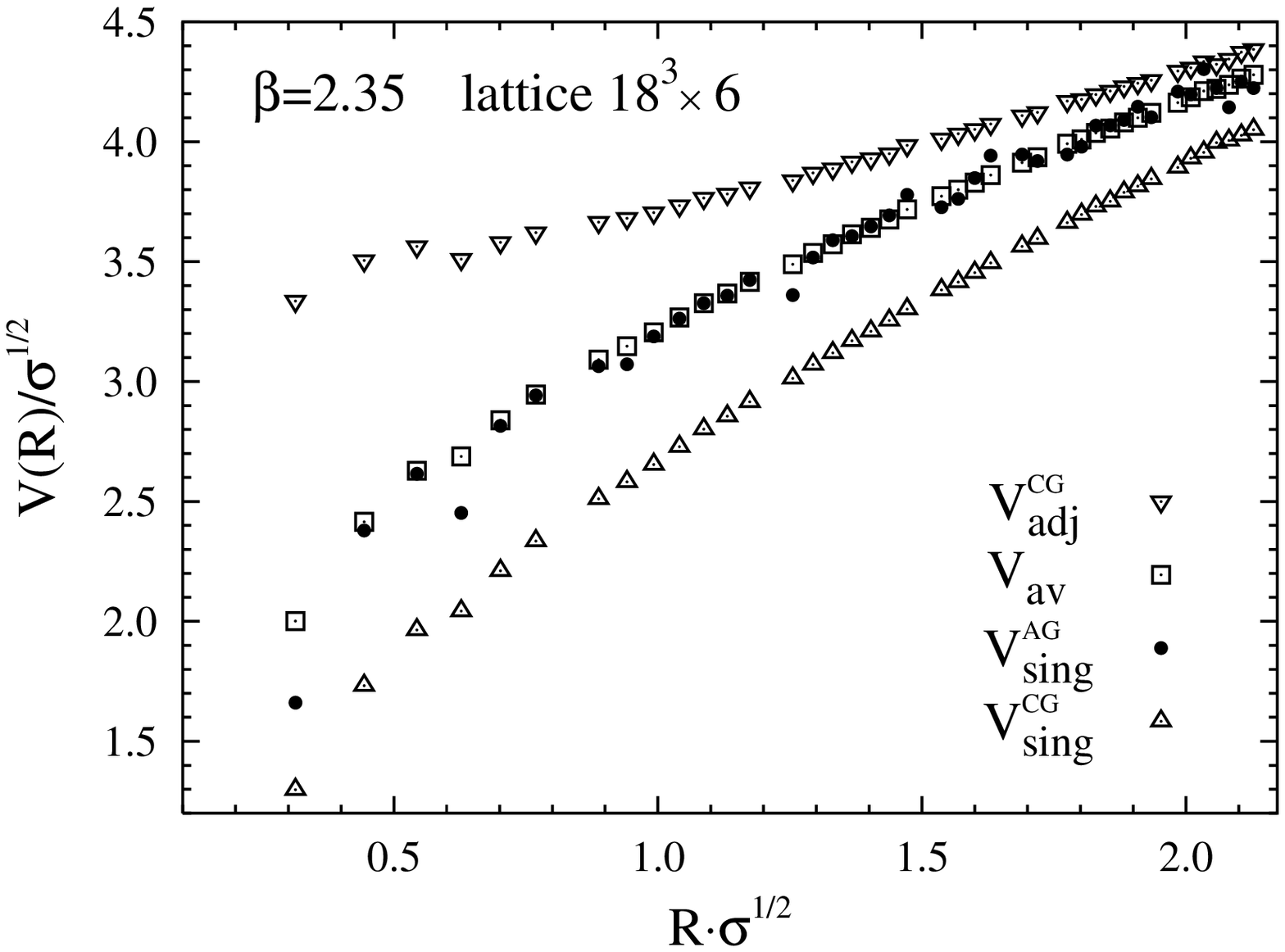}
}
\end{center}
\caption{Dependence on $R=|\vR|$ of color averaged potential
$~V_{av}(\vR)~$, singlet and adjoint potentials
$~V_{sing}^{CG}(\vR)~$,$~V_{adj}^{CG}(\vR)~$ in the Coulomb gauge and
singlet potential $~V_{sing}^{AG}(\vR)~$ in the maximal tree axial gauge.
Errorbars are smaller than symbol sizes.
}
\label{fig:fig1}
\end{figure}

In the Coulomb gauge our results are in agreement with earlier results
\cite{Digal:2003jc} for the $4d$ case. The singlet potential
$~V_{sing}^{CG}(\vR)~$ is below the color averaged $~V_{av}(\vR)~$ at
small distances and approaches it from below with increasing $~|\vR|~$,
while the adjoint potential $~V_{adj}^{CG}(\vR)~$ is above
$~V_{av}(\vR)~$  at small distances and approaches it from above at
large distances (see Figure~\ref{fig:fig1}).

Of special interest for us is the comparison of the singlet and adjoint
potentials in various time local gauges. Figure~\ref{fig:fig1} shows
the behavior of the singlet potential $~V_{sing}^{AG}(\vR)~$ in the
maximal tree axial gauge (filled circles). The important observation is
that this potential is far from being equal to the singlet potential
$~V_{sing}^{CG}(\vR)~$ defined in the Coulomb gauge (triangles up
Figure~\ref{fig:fig1}). In fact, this observation represents one of the
main results of our work.

Numerically, the potential $~V_{sing}^{AG}(\vR)~$ is very close to the
color averaged potential $~V_{sing}^{CG}(\vR)~$ for most values of the
distance $~R~$. At the same time at distances along the lattice axes
one can see clear deviations from $~V_{av}(\vR)~$ which are due to the
lack of the rotational invariance of maximal tree axial gauge. The
effects of this lack of the rotational invariance look even more
impressive for the potentials measured along the axes in the $1$st,
$2$nd and $3$rd directions separately (we do not show it in the
Figure).

In this gauge the correlator eq.(\ref{singl}) can be explicitly
expressed as a linear combination of the periodic Wilson loops (PWL),
introduced in \cite{Nadkarni:1986as}. The path connecting points
$\vzero$ and $\vR$ (space-like leg of PWL) depends on direction, e.g.
the path is a straight line for $\vR=\{R,0,0\}$, while it deviates from
a straight line for $\vR=\{0,R,0\}$ and $\vR=\{0,0,R\}$. Thus the lack
of the rotational invariance can be ascribed to the strong dependence
of  the PWL on this path.  Let us also mention that the correlator
eq.(\ref{singl}) in AG is exactly the same in the maximal tree gauge,
which includes all four directions.

In the case of another gauge we used -- GCG -- the gauge fixing
functional defined in eq.(\ref{funcGCG}) depends on the parameter
$~\lambda~$ chosen to be $~0<\lambda \le 1~$. Evidently, the choice
$~\lambda=1~$ corresponds to the standard Coulomb gauge. We found that
in this gauge both potentials, i.e the singlet potential
$~V^{\lambda}_{sing}(\vR)~$ and adjoint potential
$~V^{\lambda}_{adj}(\vR)~$, demonstrate a nontrivial dependence on
$~\lambda~$.

\begin{figure}[pt]
\begin{center}
\leavevmode
\hbox{
\epsfysize=11cm
\epsfxsize=14cm
\epsfbox{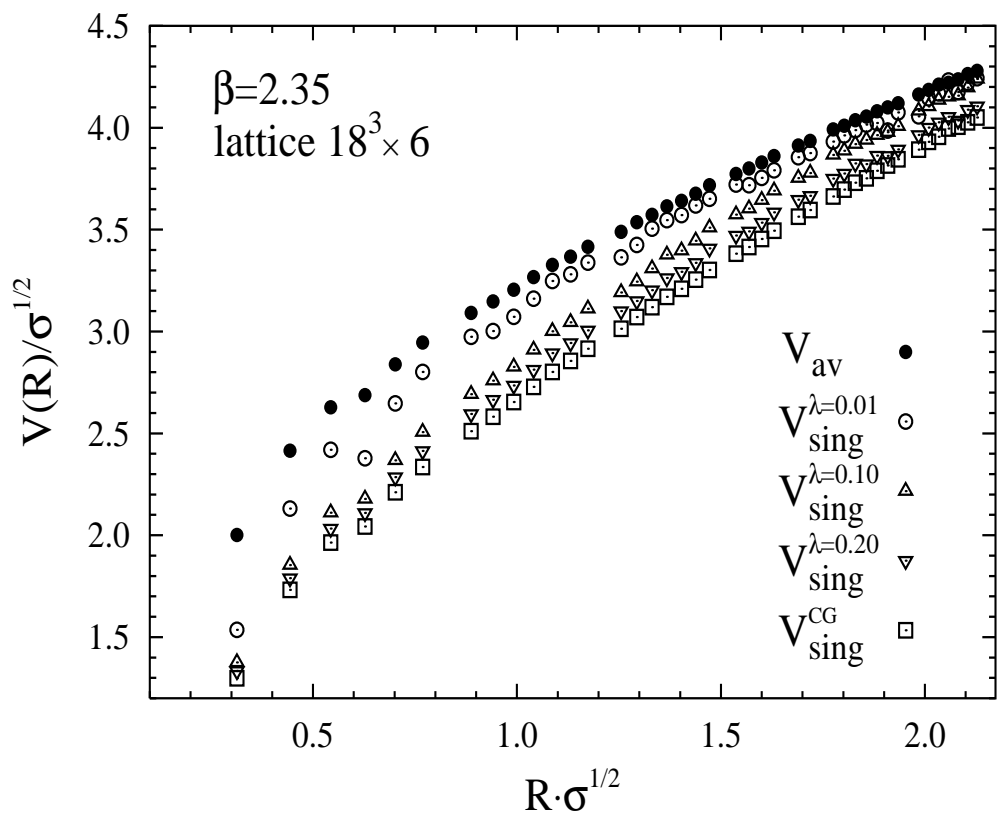}
}
\end{center}
\caption{$~R$--dependence of $~V_{sing}^{\lambda}(\vR)~$,
$~V_{sing}^{CG}(\vR)~$ and $~V_{av}(\vR)~$.
Errorbars are smaller than symbol sizes.
}
\label{fig:fig2}
\end{figure}

In Figure~\ref{fig:fig2} we show the $~R$--dependence of the singlet
potential $~V_{sing}^{\lambda}(\vR)~$ in GCG at some comparatively
small values of $~\lambda~$ as well as $~R$--dependence of
$~V_{sing}^{CG}(\vR)~$ ($\lambda=1$) and $~V_{av}(\vR)~$. One can see
that with decreasing $~\lambda~$ the singlet potential
$~V_{sing}^{\lambda}(\vR)~$ moves towards the color averaged potential
$~V_{av}(\vR)~$.

\begin{figure}[pt]
\begin{center}
\leavevmode
\hbox{
\epsfysize=11cm
\epsfxsize=14cm
\epsfbox{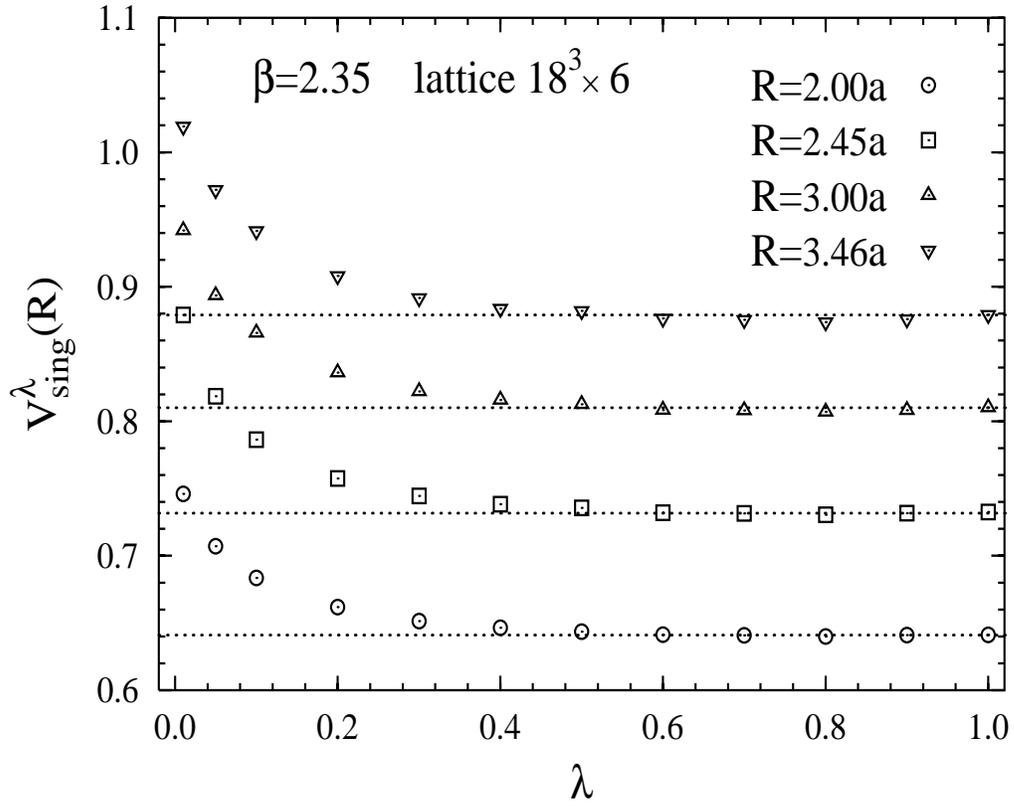}
}
\end{center}
\caption{The dependence of the singlet potential
$V^{\lambda}_{sing}(\vR)$ in GCG on $\lambda$ at various values of
the distance $R$. Errorbars are smaller than symbol sizes.
Lines correspond to the standard Coulomb gauge ($\lambda=1$).
}
\label{fig:fig3}
\end{figure}

Figure~\ref{fig:fig3} shows the dependence on $~\lambda$ of the singlet
potential $~V_{sing}^{\lambda}(\vR)~$ at some particular values of
$~R~$ and reveals a nontrivial character of this dependence. Indeed,
the data suggest that for $~\lambda \ge \lambda_c~$ ($\lambda_c\simeq
0.6$) there is no any sizable dependence on $~\lambda~$ and, therefore,
$~V_{sing}^{\lambda}(\vR) = V_{sing}^{CG}(\vR)~$ within our errorbars.
At $~\lambda < \lambda_c~$  this is not true anymore and the deviation
of $~V_{sing}^{\lambda}(\vR)~$ from $~V_{sing}^{CG}(\vR)~$ increases
with decreasing $~\lambda~$. Thus we find the class of the gauges,
namely the GCG for $~\lambda \ge \lambda_c~$, which give rise to the
same singlet potential as the Coulomb gauge does. On the other hand, we
find further evidence for the gauge dependence of the correlator
eq.(\ref{singl}).

It is worthwhile to note that the "transition" at $~\lambda =
\lambda_c~$ has nothing to do with the restoration of the rotational
invariance. Indeed, we found that at $~\lambda > \lambda_c~$ the
singlet potential measured along the $1$st direction is different from
the potentials measured along the $2$nd or $3$rd directions.

At the end we want to make a remark about the calculation of the
singlet potential at $T=0$ using gauge noninvariant correlators of the
Wilson lines. It was demonstrated in \cite{Philipsen:2002az} that the
singlet potential calculated from such correlators in the local in time
Laplacian gauge was in full agreement with the gauge independent
calculation using Wilson loops \footnote{The correlator of the Wilson
lines summed over spatial 2-plane was used to calculate the string
tension in SU(3) theory in \cite{Marinari:1992kh}.}. We have calculated
the correlator of the Wilson lines in GCG at $\lambda=0.2$ and in CG on
$12^4$ lattice at $\beta=2.3$. We found that the singlet potentials in
both gauges agreed nicely with the potential determined from the Wilson
loops. Thus at $T=0$ the gauge dependent correlators of the Wilson
lines give rise to the same singlet potentials in GCG as in the
Laplacian gauge or in CG. In AG the noise was too strong to make any
definite conclusion.

We conclude that the statement made in \cite{Philipsen:2002az} about
uniqueness of the singlet and adjoint potentials defined by
eq.'s~(\ref{singl}),(\ref{adjo}) in any unique and local in time gauge
is possibly too strong. We believe that further study of the definition
of the singlet and adjoint potentials at finite temperature will be
useful. Our study can be extended to other, rotationally invariant
gauges, different from the Laplacian gauge and the Coulomb gauge. An
example of such gauge is the 3D Maximal Abelian gauge
\cite{Chernodub:1998aj} with additional fixing of the abelian degrees
of freedom or its analog using corresponding Laplacian operators.

\vspace{5mm}
This work has been supported by the grant INTAS-00-00111, RFBR grant
02-02-17308 and Heisenberg--Landau program.

\newpage


%

\vfill


%

\vfill


%

\vfill

\end{document}